\documentclass[12pt]{article}
\usepackage{amsmath}
\def\a{\alpha}
\def\b{\beta}

\def\d{\delta}
\def\e{\epsilon}

\def\g{\gamma}
\def\p{\psi}

\def\t{\theta}
\def\r{\rho}
\def\s{\sigma}

\def\be{\begin{equation}}
\def\ee{\end{equation}}
\def\arr{\begin{array}{rll}}
\def\ea{\end{array}}
\def\bea{\begin{eqnarray}}
\def\eea{\end{eqnarray}}

\def\N2{$N{=}2$}

\def\>{\rangle}
\def\<{\langle}
\def\+{\dagger}
\def\={\ =\ }

\begin{document}
\renewcommand{\thefootnote}{\fnsymbol{footnote}}
\begin{titlepage}
\setcounter{page}{0}
\vskip 1cm
\begin{center}
{\LARGE\bf  Couplings in $D(2,1;\alpha)$ superconformal    }\\
\vskip 0.5cm
{\LARGE\bf mechanics from the $SU(2)$ perspective}\\
\vskip 2cm
$
\textrm{\Large Anton Galajinsky\ }
$
\vskip 0.7cm
{\it
Laboratory of Mathematical Physics, Tomsk Polytechnic University, \\
634050 Tomsk, Lenin Ave. 30, Russian Federation} \\
{Email: galajin@tpu.ru}
\vskip 0.5cm

\end{center}
\vskip 1cm
\begin{abstract} \noindent
Dynamical realizations of the most general $N=4$ superconformal group in one dimension $D(2,1;\alpha)$ are reconsidered from the perspective of the $R$--symmetry subgroup $SU(2)$.
It is shown that any realization of the $R$--symmetry subalgebra in some phase space can be extended to a representation of the Lie superalgebra corresponding to $D(2,1;\alpha)$.
Novel couplings of arbitrary number of supermultiplets of the type $(1,4,3)$ and $(0,4,4)$ with a single supermultiplet of either the type $(3,4,1)$, or $(4,4,0)$ are constructed.
$D(2,1;\alpha)$ superconformal mechanics describing superparticles propagating near the horizon of the extreme Reissner--Nordstr\"om--AdS--dS black hole in four and five dimensions is considered.
The parameter $\alpha$ is linked to the cosmological constant.
\end{abstract}

\vspace{0.5cm}

PACS: 11.30.Pb; 12.60.Jv; 04.70.Bw\\ \indent
Keywords: $D(2,1;\alpha)$ superconformal group, near horizon black holes
\end{titlepage}

\renewcommand{\thefootnote}{\arabic{footnote}}
\setcounter{footnote}0

\noindent
{\bf 1. Introduction}\\

The exceptional supergroup $D(2,1;\alpha)$ describes the most general $N=4$ supersymmetric extension of the conformal group in one dimension $SO(2,1)$. It is parametrized by one real number $\alpha$. As far as realizations in superspace are concerned, the generators of the corresponding Lie superalgebra are associated with time translations, dilatations, special conformal transformations, supersymmetry transformations and their superconformal partners, as well as with two variants of $su(2)$--transformations. One $su(2)$ is interpreted as the $R$--symmetry subalgebra, while the other affects only fermions. Recent interest in $D(2,1;\alpha)$ and specifically in $SU(1,1|2)$ which arises at $\alpha=-1$ was  motivated by a possible link to a microscopic description of the near horizon extreme Reissner--Nordstr\"om black hole and the desire to better understand peculiar features of extended supersymmetry in $d=1$ which are absent in higher dimensions\footnote{The literature on the subject is rather extensive. For recent developments and further references see \cite{G,IF}.}. It is curious that none of the $D(2,1;\alpha)$ superconformal mechanics models considered thus far assigned any physical meaning to the parameter $\alpha$ (for geometric interpretations see \cite{FIL2}).

A related line of research is the construction of superconformal particles propagating on near horizon extreme black hole backgrounds. Such systems can be linked to the conventional superconformal mechanics by applying a proper coordinate transformation \cite{IKN,BGIK}. It is believed that they will help to establish a precise relation between supergravity Killing spinors and supersymmetry charges of superparticles on curved backgrounds.

In a recent work \cite{G}, couplings in $SU(1,1|2)$ superconformal mechanics have been reconsidered  from the perspective of the $R$--symmetry subgroup $SU(2)$. It was shown that
any realization of $su(2)$ in terms of phase space functions can be extended to a representation of the full $su(1,1|2)$ superconformal algebra. Novel interactions of supermultiplet of the type $(0,4,4)$ to either $(1,4,3)$--, or $(3,4,1)$--, or $(4,4,0)$--supermultiplet have been constructed by arranging the $su(2)$--generators so as to include both bosons and fermions.
Soon after, in Ref. \cite{IF} the off--shell superfield method has been applied so as to generalize the results in \cite{G} to the case of $D(2,1;\alpha)$ and even more general coupling involving three supermultiplets of the type $(1,4,3)$, $(4,4,0)$, and $(0,4,4)$ has been built.

Conventional means of building superconformal mechanics include the superfield approach, the method of nonlinear realizations, and the direct construction of a representation of the desired
superconformal algebra within the Hamiltonian framework. While the superfield technique is definitely more powerful, the Hamiltonian approach is more efficient in analyzing the dynamical content and the structure of interactions because non--dynamical auxiliary fields are absent.

The goal of this work is to extend the Hamiltonian analysis in \cite{G} to the case of the exceptional supergroup $D(2,1;\alpha)$. In doing so, we recover the results in a recent work \cite{IF} and further extend them by constructing a $D(2,1;\alpha)$--invariant model which describes coupling of arbitrary number of supermultiplets of the type $(1,4,3)$ and $(0,4,4)$ to a single supermultiplet of either the type $(3,4,1)$, or $(4,4,0)$. We also discuss $D(2,1;\alpha)$ superconformal mechanics in the so called AdS basis \cite{IKN} and connect the systems based on $(3,4,1)$--, and $(4,4,0)$--supermultiplets to superparticles propagating near the horizon of the extreme Reissner--Nordstr\"om--AdS--dS black hole in four and five dimensions. In that context, the parameter $\alpha$ is linked to the cosmological constant and thus, for the first time in the literature, it is given a clear physical interpretation.

The work is organized as follows. In the next section it is argued that any representation of the $R$--symmetry subalgebra $su(2)$ in terms of phase space functions can be automatically extended to a representation of the Lie superalgebra associated with $D(2,1;\alpha)$. In Sect. 3, based on the earlier work \cite{KL}, we construct a novel coupling of an arbitrary number of supermultiplets of the type $(1,4,3)$ and $(0,4,4)$ to a single supermultiplet of either the type $(3,4,1)$, or $(4,4,0)$. $D(2,1;\alpha)$--superparticles on black hole backgrounds are considered in Sect. 4. Models associated with the near horizon geometry of the extreme Reissner--Nordstr\"om--AdS--dS black hole in four and five dimensions are linked to $D(2,1;\alpha)$ superconformal mechanics based on supermultiplets of the type $(3,4,1)$ and $(4,4,0)$, respectively. The parameter $\alpha$ is linked to the cosmological constant. Our spinor conventions are gathered in Appendix. Throughout the paper summation over repeated indices is understood.

\vspace{0.5cm}

\noindent
{\bf 2. Extending $su(2)$ to $D(2,1;\alpha)$}\\

Consider a representation of $su(2)$ in terms of functions on some phase space
\be\label{su(2)}
\{J_a,J_b\}=\e_{abc} J_c,
\ee
where $a=1,2,3$ and $\e_{abc}$ is the totally antisymmetric symbol with $\epsilon_{123}=1$. In what follows three realizations
we will of interest. The first is given by the angular momentum of a free particle moving on two--dimensional sphere
\bea\label{S2}
&&
J_1=-p_\Phi \cot \Theta \cos \Phi  -p_\Theta \sin \Phi,
\nonumber\\[2pt]
&&
J_2=-p_\Phi \cot \Theta \sin \Phi +p_\Theta \cos \Phi,
\nonumber\\[2pt]
&&
J_3=p_\Phi, \quad J_a J_a=p_\Theta^2+p_\Phi^2 \sin^{-2}\Theta,
\eea
where $(\Theta,p_\Theta)$ and $(\Phi,p_\Phi)$ form canonical pairs obeying the conventional Poisson brackets $\{\Theta,p_\Theta \}=1$, $\{\Phi,p_\Phi \}=1$. The second is provided by the same model in external field of the Dirac monopole
\bea\label{s2}
&&
J_1=-p_\Phi \cot \Theta \cos \Phi  -p_\Theta \sin \Phi+q \cos \Phi \sin^{-1}\Theta,
\nonumber\\[2pt]
&&
J_2=-p_\Phi \cot \Theta \sin \Phi +p_\Theta \cos \Phi +q \sin \Phi \sin^{-1}\Theta,
\nonumber\\[2pt]
&&
J_3=p_\Phi, \quad J_a J_a=p_\Theta^2+{(p_\Phi-q \cos\Theta)}^2 \sin^{-2}\Theta  +q^2,
\eea
where $q$ is the magnetic charge. The third is linked to the geodesic motion on the group manifold $SU(2)$ and is given by the vector fields dual to the conventional left--invariant one--forms
\bea\label{s3}
&&
J_1=-p_\Phi \cot \Theta \cos \Phi  -p_\Theta \sin \Phi+p_\Psi \cos \Phi \sin^{-1}\Theta,
\nonumber\\[2pt]
&&
J_2=-p_\Phi \cot \Theta \sin \Phi +p_\Theta \cos \Phi +p_\Psi \sin \Phi \sin^{-1}\Theta,
\nonumber\\[2pt]
&&
J_3=p_\Phi, \quad J_a J_a=p_\Theta^2+{(p_\Phi-p_\Psi \cos\Theta)}^2 \sin^{-2}\Theta +p_\Psi^2,
\eea
with $(\Theta,p_\Theta)$, $(\Phi,p_\Phi)$ and $(\Psi,p_\Psi)$ forming the canonical pairs.

Note that (\ref{s2}) follows from (\ref{S2}) by introducing the coupling to the external vector field potential $p_a \rightarrow p_a+A_a (\Theta,\Phi)$
and imposing the structure relations of $su(2)$, while (\ref{s3}) results from (\ref{s2}) by implementing the oxidation $q \rightarrow p_\Psi$ with respect to the constant $q$.
Focusing on the Casimir element $J_a J_a$, it is important to stress that
all the $su(2)$--realizations exhibited above are characterized by a non--degenerate metric which accompanies terms quadratic in momenta.  Direct sums of $J_a$ in (\ref{S2}), (\ref{s2}), (\ref{s3}) yield degenerate metrics which prove to be unsuitable for the applications to follow.

Each realization of $su(2)$ in a phase space can be extended to a representation of the Lie superalgebra corresponding to $D(2,1;\alpha)$.
It suffices to introduce an extra bosonic canonical pair $(x,p)$ along with a fermionic $SU(2)$--spinor $\psi_\alpha$, $\alpha=1,2$, and its complex conjugate ${(\psi_\alpha)}^{*}=\bar\p^\a$, and impose the brackets\footnote{Within the Hamiltonian formalism the canonical bracket $\{ \p_\a, \bar\p^\b \}=-i{\d_\a}^\b$ is conventionally understood as
the Dirac bracket $
{\{A,B \}}_D=\{A,B \}-i\{A,\lambda^\a \}\{\bar\lambda_\a,B \}-i\{A,\bar\lambda_\a \}\{\lambda^\a,B \}$
associated with the fermionic second class constraints
$\lambda^\a={p_\p}^\a-\frac i2 \bar\p^\a=0$ and $\bar\lambda_\a=p_{\bar\p \a}-\frac i2 \p_\a=0$. Here
$({p_\p}^\a,p_{\bar\p \a})$ stand for the momenta canonically conjugate to the variables
$(\p_\a,\bar\p^\a)$, respectively. Choosing the right derivative for the fermionic degrees of freedom, the action functional, which reproduces the Dirac bracket for the fermionic pair, reads
$S=\int dt \left(\frac i2 \bar\p^\a {\dot\p}_\a-\frac i2 {\dot{\bar\p}} {}^\a \p_\a\right)$. Similar consideration applies to the fermionic pair $(\chi_\alpha,\bar\chi^\alpha)$ which appears in Sect. 3.
}
\be\label{cr}
\{x,p\}=1\ , \qquad \{ \p_\a, \bar\p^\b \}=-i{\d_\a}^\b.
\ee
Then it is straightforward to verify that the functions
\begin{align}\label{rep}
&
H=\frac{p^2}{2}+\frac{2 \alpha^2 }{x^2} J_a J_a+\frac{2 \alpha}{x^2} (\bar\p \s_a \p) J_a -\frac{(1+2\alpha)}{4x^2} \p^2 \bar\p^2, && D=tH-\frac 12 x p,
\nonumber\\[2pt]
&
K=t^2 H-t x p +\frac 12 x^2, && \mathcal{J}_a=J_a+\frac 12 (\bar\p \s_a \p),
\nonumber\\[2pt]
&
Q_\a=p \p_\a-\frac{2i\alpha}{x} {(\s_a \p)}_\a J_a -\frac{i(1+2\alpha)}{2x} \bar\p_\a \p^2\ , && S_\a=x \p_\a -t Q_\a,
\nonumber\\[2pt]
&
\bar Q^\a =p \bar\p^\a+\frac{2i\alpha}{x} {(\bar\p \s_a)}^\a J_a -\frac{i(1+2\alpha)}{2x} \p^\a \bar\p^2, &&
\bar S^\a=x \bar\p^\a -t \bar Q^\a,
\nonumber\\[2pt]
&
I_{-}=\frac{i}{2} \psi^2, \qquad \qquad \qquad \qquad I_{+}=-\frac{i}{2} {\bar\psi}^2, &&
I_3=\frac 12 \bar\psi \psi,
\end{align}
where $\s_a$ are the Pauli matrices (for our spinor conventions see Appendix),
do obey the structure relations of the Lie superalgebra corresponding to $D(2,1;\alpha)$
\begin{align}\label{algebra}
&
\{ H,D \}=H, && \{ H,K \}=2D,
\nonumber\\[2pt]
&
\{D,K\}=K, && \{ \mathcal{J}_a,\mathcal{J}_b \}=\epsilon_{abc} \mathcal{J}_c,
\nonumber\\[2pt]
&
\{ Q_\a, \bar Q^\b \}=-2 i H {\d_\a}^\b, &&
\{ Q_\a, \bar S^\b \}=-2\alpha {{(\s_a)}_\a}^\b \mathcal{J}_a+2iD {\d_\a}^\b+2(1+\alpha)I_3 {\d_\a}^\b,
\nonumber\\[2pt]
&
\{ S_\a, \bar S^\b \}=-2i K {\d_\a}^\b, &&
\{ \bar Q^\a, S_\b \}=2\alpha{{(\s_a)}_\b}^\a \mathcal{J}_a+2iD {\d_\b}^\a-2(1+\alpha)I_3 {\d_\b}^\a,
\nonumber\\[2pt]
&
\{ Q_\a, S_\b \}=2i (1+\alpha) \epsilon_{\alpha \beta} I_{-}, &&
\{ {\bar Q}^\a, {\bar S}^\b \}=-2i (1+\alpha) \epsilon^{\alpha \beta} I_{+},
\nonumber\\[2pt]
& \{ D,Q_\a\} = -\frac{1}{2} Q_\a, && \{ D,S_\a\} =\frac{1}{2} S_\a,
\nonumber\\[2pt]
&
\{ K,Q_\a \} =S_\a, && \{ H,S_\a \}=-Q_\a,
\nonumber\\[2pt]
&
\{ \mathcal{J}_a,Q_\a\} =\frac{i}{2} {{(\s_a)}_\a}^\b Q_\b, && \{ \mathcal{J}_a,S_\a\} =\frac{i}{2} {{(\s_a)}_\a}^\b S_\b,
\nonumber\\[2pt]
& \{ D,\bar Q^\a \} =-\frac{1}{2} \bar Q^\a, && \{ D,\bar S^\a\} =\frac{1}{2} \bar S^\a,
\nonumber\\[2pt]
& \{K,\bar Q^\a\} =\bar S^\a, && \{ H,\bar S^\a\} =-\bar Q^\a,
\nonumber\\[2pt]
&
\{\mathcal{J}_a,\bar Q^\a\} =-\frac{i}{2} \bar Q^\b {{(\s_a)}_\b}^\a, && \{ \mathcal{J}_a,\bar S^\a\} =-\frac{i}{2}
\bar S^\b {{(\s_a)}_\b}^\a,
\nonumber\\[2pt]
&
\{ I_{-},\bar Q^\a \} =\epsilon^{\alpha \beta} Q_\beta, && \{ I_{-},\bar S^\a \} =\epsilon^{\alpha \beta} S_\beta,
\nonumber\\[2pt]
&
\{ I_{+},Q_\a\} =-\epsilon_{\alpha\beta} \bar Q^\b, && \{ I_{+},S_\a\} =-\epsilon_{\alpha\beta} \bar S^\b,
\nonumber\\[2pt]
&
\{ I_3,Q_\a \} =\frac{i}{2} Q_\a, &&  \{ I_3,S_\a \} =\frac{i}{2} S_\a,
\nonumber\\[2pt]
&
\{ I_3,\bar Q^\a\} =-\frac{i}{2} \bar Q^\a, &&  \{ I_3,\bar S^\a \} =-\frac{i}{2} \bar S^\a,
\nonumber\\[2pt]
&
\{ I_{-},I_3\} =-i I_{-}, &&  \{ I_{+},I_3 \} =i I_{+},
\nonumber\\[2pt]
&
\{ I_{-},I_{+}\} =2 i I_{3}. &&
\end{align}
When verifying the structure relations (\ref{algebra}), the properties of the Pauli matrices and the spinor identities gathered in Appendix were extensively used.

As far as dynamical realizations are concerned, $H$ is interpreted as the Hamiltonian. $D$ and $K$ are treated as the generators of dilatations and special conformal transformations. $Q_\a$ are the supersymmetry generators and $S_\a$ are their superconformal partners. $\mathcal{J}_a$ generate the $R$--symmetry subalgebra $su(2)$. So do also $I_{\pm}$, $I_3$ for which the Cartan basis is chosen.
The extra $su(2)$, which is realized on the fermions, makes the main difference with the $su(1,1|2)$ superconformal algebra, which arises at $\alpha=-1$.

It should be mentioned that a representation similar to (\ref{rep}) was first considered in \cite{FIL2}. Yet, the functions $J_a$ were assigned quite a different meaning. In \cite{FIL2} they involved non--dynamical harmonic variables which represented spin degrees of freedom. In this work, we suggest to realize $J_a$ in terms of the fully fledged dynamical variables as displayed in Eqs. (\ref{S2}), (\ref{s2}), (\ref{s3}) above. Then Eqs. (\ref{rep}) provide a Hamiltonian description of $D(2,1;\alpha)$--supermultiples of the type $(3,4,1)$ (two on-shell versions) or $(4,4,0)$. Worth mentioning is also the work in \cite{HKLN} where it was demonstrated that the angular part of a generic conformal mechanics can be lifted to a $D(2,1;\alpha)$--invariant system.

\vspace{0.5cm}

\noindent
{\bf 3. Couplings of $D(2,1;\alpha)$ supermultiplets from the $su(2)$ perspective  }\\

\noindent

In a recent work \cite{G}, we reexamined dynamical realizations of the superconformal algebra $su(1,1|2)$ and constructed novel on-shell couplings of
supermultiplet of the type $(0,4,4)$ to a single supermultiplet of either the type $(1,4,3)$, or $(3,4,1)$, or $(4,4,0)$. This was achieved by
introducing an extra pair of complex conjugate fermions $\chi_\alpha$,
$\bar\chi^\a={(\chi_\alpha)}^{*}$, $\alpha=1,2$, which obey the canonical bracket
\be
\{ \chi_\a, \bar\chi^\b \}=-i{\d_\a}^\b,
\ee
and promoting $J_a$ in (\ref{su(2)}) to
\be\label{s4}
\tilde J_a=J_a+\frac 12 (\bar\chi \sigma_a \chi).
\ee
While one cannot consistently combine two bosonic realizations of $su(2)$ within an unconstrained dynamical system with $D(2,1;\alpha)$--superconformal symmetry, the direct sum of $J_a$ in (\ref{S2}), or (\ref{s2}), or (\ref{s3}) with $\frac 12 (\bar\chi \sigma_a \chi)$ is admissible. All one needs in verifying the structure relations of the Lie superalgebra corresponding to $D(2,1;\alpha)$ is that $\tilde J_a$ form $su(2)$.
The resulting (on--shell) model can be viewed as describing a particular interaction of $(0,4,4)$--supermultiplet realized on the pair $(\chi_\alpha,\bar\chi^\a)$ with either $(3,4,1)$--, or $(4,4,0)$--supermultiplet.
The corresponding off--shell superfield Lagrangian formulations have been constructed in \cite{IF} which generalize the on-shell $su(1,1|2)$--superconformal models in \cite{G}. Yet, in \cite{IF} it was also shown that the superfield approach is capable of describing a similar interaction between three distinct supermultiplets of the type $(1,4,3)$, $(4,4,0)$, and $(0,4,4)$. Below we generalize that result by coupling an arbitrary number of $(1,4,3)$--, and $(0,4,4)$--supermultiples to a single supermultiplet of either the type $(3,4,1)$, or $(4,4,0)$. The idea is to build a many--body generalization of the representation (\ref{rep}) in the spirit of \cite{GLP}.

Consider a set of canonical pairs which involves bosons $(x^i,p^i)$ and fermions $(\psi^i_\alpha,\bar\p^{i \a})$, $(\chi^A_\alpha,\bar\chi^{A \a})$, with $i=1,\dots,M+1$, $A=1,\dots, N$, $\alpha=1,2$, obeying the brackets
\be\label{cr}
\{x^i,p^j\}=\delta^{ij}, \qquad \{ \p^i_\a, \bar\p^{j \b} \}=-i \delta^{ij}{\d_\a}^\b, \qquad  \{ \chi^A_\a, \bar\chi^{B \b} \}=-i \delta^{AB}{\d_\a}^\b.
\ee
Guided by our previous study of the $su(1,1|2)$ superconformal mechanics \cite{GLP}, on such a phase space we introduce ansatze for the $D(2,1;\alpha)$--generators
\begin{align}
&
H=\frac{1}{2} p^i p^i+\frac 12 \partial^i V \partial^i V \tilde J_a \tilde J_a+\partial^i \partial^j V \tilde J_a (\bar\p^i \s_a \p^j)  -\frac 12 \partial^i W^{jkl} (\psi^i \psi^j) (\bar\psi^k \bar\psi^l), && D=tH-\frac 12 x^i p^i,
\nonumber\\[2pt]
&
K=t^2 H-t x^i p^i +\frac 12 x^i x^i, && I_3=\frac 12 (\bar\psi^i \psi^i),
\nonumber\\[2pt]
&
I_{-}=\frac{i}{2} (\psi^i \psi^i), && I_{+}=-\frac{i}{2} ({\bar\psi}^i {\bar\psi}^i),
\nonumber
\end{align}
\begin{align}\label{repre}
&
Q_\a=p^i \p^i_\a+i \partial^i V {(\s_a \p^i)}_\a \tilde J_a +i W^{ijk} \bar\p^i_\a (\psi^j \psi^k)\ , && S_\a=x^i \p^i_\a -t Q_\a,
\nonumber\\[2pt]
&
\bar Q^\a =p^i \bar\p^{i \a}-i \partial^i V {(\bar\p^i \s_a)}^\a \tilde J_a +i W^{ijk} \p^{i \a} (\bar\psi^j \bar\psi^k), &&
\bar S^\a=x^i \bar\p^{i \a} -t \bar Q^\a,
\nonumber\\[2pt]
&
\mathcal{J}_a=\tilde J_a+\frac 12 (\bar\p^i \s_a \p^i), &&
\end{align}
which involve two scalar prepotentials $V=V(x)$, $F=F(x)$ with $W^{ijk}=\partial^i \partial^j \partial^k F(x)$ and $\tilde J_a=J_a+\frac 12 (\bar\chi^A \sigma_a \chi^A)$. It is assumed that $J_a$ is one of the $su(2)$ realizations exposed in Eqs. (\ref{S2}), (\ref{s2}), (\ref{s3}) above. The structure relations (\ref{algebra}) impose the following constraints on the prepotentials:
\begin{align}\label{strr}
&
\partial^i \partial^j V+\partial^i V \partial^j V-2 W^{ijk} \partial^k V=0, && W^{ijk} W^{klm}=W^{mjk} W^{kli},
\nonumber\\[2pt]
&
x^i \partial^i V=-2 \alpha, && x^i W^{ijk}=-\frac 12 (1+2\alpha) \delta^{jk}.
\end{align}
When verifying (\ref{algebra}), the spinor algebra and the properties of the Pauli matrices given in Appendix were extensively used. Note that for a nonzero value of the parameter $\alpha$ the system (\ref{strr}) does not allow the prepotential $V$ to vanish. The restrictions (\ref{strr}) have been obtained under the assumption that the $su(2)$ generators $J_a$ in Eqs. (\ref{s4}), (\ref{repre}) are nontrivial. The choice $J_a=0$, which is also compatible with (\ref{su(2)}), would have altered the leftmost equation entering the first line in (\ref{strr}) \cite{KL,GLP}.

Inspired by the earlier work \cite{FIL2}, a representation similar to (\ref{repre}) has been constructed in \cite{KL}~\footnote{The constraints (\ref{strr}) fit those in \cite{KL} after the redefinition $V \to -V$, $2 W^{ijk} \to W^{ijk}$.}. In particular, a plenty of interesting solutions to the master equations
(\ref{strr}) have been found, which relied upon the root systems and their deformations. Yet, like in \cite{FIL2}, the functions $\tilde J_a$ were realized in terms of non--dynamical spin degrees of freedom and the possibility to include into the consideration $N$ copies of
$(0,4,4)$--supermultiplet described by $(\chi^A_\alpha,\bar\chi^{A \a})$ remained unnoticed.

Being combined with the solutions to Eq. (\ref{strr}) in \cite{KL}, the representation (\ref{repre}) gives a clue to building novel couplings in $D(2,1;\alpha)$ superconformal mechanics.  Given a particular solution to (\ref{strr}), the resulting model (\ref{repre}) describes an interaction of $M$ supermultiplets of the type $(1,4,3)$ with $N$ $(0,4,4)$--supermultiplets and a single supermultiplet of either the type $(3,4,1)$, or $(4,4,0)$. Alternatively, one can regard this system as describing a coupling of $M+1$ copies of either $(3,4,1)$--, or $(4,4,0)$--supermultiplet, in which angular degrees of freedom are identified, to $N$ supermultiplets of the type $(0,4,4)$.

Completing this section, we exhibit the on--shell Lagrangian formulations associated with the Hamiltonian description (\ref{repre}) and assume $\alpha\ne 0$ which excludes $V=0$.
Given $J_a$ in (\ref{s2}), let us introduce the $3$--vector $\lambda_a$ parameterizing a point on the unit sphere
\be
\lambda_a=(\cos{\Phi} \sin{\Theta}, \sin{\Phi} \sin{\Theta},\cos{\Theta}), \qquad \lambda_a \lambda_a=1,
\ee
and the $3$--vector $\mathcal{L}_a$
\be
\mathcal{L}_a=\dot\Theta \left(\frac{\partial J_a}{\partial p_\Theta}\right)+\dot \Phi \sin^2{\Theta} \left(\frac{\partial J_a}{\partial p_\Phi}\right)+q \partial^i V \partial^i V  \lambda_a,  \qquad  \mathcal{L}_a \lambda_a=q \partial^i V \partial^i V,
\ee
which has the components
\bea\label{LL}
&&
\mathcal{L}_1=-\dot\Theta \sin{\Phi} -\dot\Phi \sin{\Theta} \cos{\Theta} \cos{\Phi}+q \partial^i V \partial^i V  \sin{\Theta} \cos{\Phi},
\nonumber\\[2pt]
&&
\mathcal{L}_2=\dot\Theta \cos{\Phi} -\dot\Phi \sin{\Theta} \cos{\Theta} \sin{\Phi}+q \partial^i V \partial^i V \sin{\Theta} \sin{\Phi},
\nonumber\\[2pt]
&&
\mathcal{L}_3=\dot\Phi \sin^2 {\Theta}+q \partial^i V \partial^i V \cos{\Theta}, \quad
\mathcal{L}_a \mathcal{L}_a={\dot\Theta}^2
+{\dot\Phi}^2 \sin^2 \Theta+q^2 {\left(\partial^i V \partial^i V \right)}^2.
\eea
Then the on--shell Lagrangian which describes a coupling of a single supermultiplet of the type $(3,4,1)$ to and arbitrary number of $(1,4,3)$--, and $(0,4,4)$--supermultiplets reads
\bea\label{action}
&&
S=\int dt \left(\frac 12 {\dot x}^i \dot x^i +\frac i2 \bar\p^i {\dot\p}^i-\frac i2 {\dot{\bar\p}} {}^i \p^i+\frac i2 \bar\chi^A {\dot\chi}^A-\frac i2 {\dot{\bar\chi}} {}^A \chi^A+\frac 12 {\left(\partial^i V \partial^i V \right)}^{-1} \left({\dot\Theta}^2
+{\dot\Phi}^2 \sin^2 \Theta \right)
\right.
\nonumber\\[2pt]
&&
\left.
\quad
-\frac{q^2}{2} \left(\partial^i V \partial^i V \right)
+ q  \dot\Phi \cos{\Theta}
-{\left(\partial^l V \partial^l V \right)}^{-1} \left(\partial^i \partial^j V (\bar\p^i \s_a \p^j)+\frac 12 \partial^i V \partial^i V (\bar\chi^A \s_a \chi^A)\right) \mathcal{L}_a
\right.
\nonumber\\[2pt]
&&
\left.
\quad
-\frac{1}{2} {\left(\partial^i V \partial^i V \right)}^{-1}  {\left[\left(\partial^i \partial^j V (\bar\p^i \s_a \p^j)+\frac 12 \partial^i V \partial^i V (\bar\chi^A \s_a \chi^A)\right)\lambda_a\right]}^2
\right.
\nonumber\\[2pt]
&&
\left.
\quad
+\left(\partial^i W^{jkl}-{ \left(\partial^p V \partial^p V \right)}^{-1} \left[\frac 12  \partial^i \partial^j V \partial^k \partial^l V +\partial^i \partial^l V \partial^k \partial^j V \right] \right) (\bar\p^i\p^j) (\bar\p^k \p^l)\right).
\eea
In obtaining (\ref{action}), one has to rewrite the phase space functions (\ref{s2}) within the Lagrangian framework
\be
J_a= { \left(\partial^i V \partial^i V \right)}^{-1}  \left(\mathcal{L}_a-B_a+(B_c \lambda_c) \lambda_a \right),
\ee
where $B_a=\left(\partial^i \partial^j V (\bar\p^i \s_a \p^j)+\frac 12 \partial^i V \partial^i V (\bar\chi^A \s_a \chi^A)\right)$. Note that the kinetic terms for the fermions correlate with the form of the canonical (Dirac) bracket chosen above (see the footnote on page 3).
Alternatively, the system (\ref{action}) can be viewed as describing an interaction of $M+1$ copies of $(3,4,1)$--supermultiplet, in which angular degrees of freedom are identified, with $N$ supermultiplets of the type $(0,4,4)$.

The Lagrangian system based on the realization of $su(2)$ in (\ref{s3}) is constructed likewise. Introducing the $3$--vector $\mathcal{L}_a$
\be
\mathcal{L}_a=\dot \Theta \left(\frac{\partial J_a}{\partial p_\Theta} \right)+\left(\dot\Phi+\dot\Psi \cos{\Theta} \right)   \left(\frac{\partial J_a}{\partial p_\Phi}\right)
+\left(\dot\Psi+\dot\Phi \cos{\Theta} \right) \left(\frac{\partial J_a}{\partial p_\Psi}\right),
\ee
which has the components
\bea\label{compL}
&&
\mathcal{L}_1=-\dot\Theta \sin{\Phi} +\dot\Psi \sin{\Theta} \cos{\Phi},
\qquad
\mathcal{L}_2=\dot\Theta \cos{\Phi} +\dot\Psi \sin{\Theta} \sin{\Phi},
\nonumber\\[2pt]
&&
\mathcal{L}_3=\dot\Phi +\dot\Psi \cos{\Theta}, \qquad ~ \mathcal{L}_a\mathcal{L}_a={\dot\Theta}^2
+{\dot\Phi}^2 \sin^2 \Theta +{(\dot\Psi+\dot\Phi \cos{\Theta} )}^2,
\eea
and implementing the inverse Legendre transformation to the Hamiltonian in (\ref{repre}), one gets
\bea
&&
S=\int dt \left(\frac 12 {\dot x}^i \dot x^i +\frac i2 \bar\p^i {\dot\p}^i-\frac i2 {\dot{\bar\p}} {}^i \p^i+\frac i2 \bar\chi^A {\dot\chi}^A-\frac i2 {\dot{\bar\chi}} {}^A \chi^A
\right.
\nonumber\\[2pt]
&&
\left.
\quad \quad
+\frac 12 {\left(\partial^i V \partial^i V \right)}^{-1} \left({\dot\Theta}^2
+{\dot\Phi}^2 \sin^2 \Theta +{(\dot\Psi+\dot\Phi \cos{\Theta} )}^2 \right)
\right.
\nonumber
\eea
\bea\label{action1}
&&
\left.
\quad
-{\left(\partial^l V \partial^l V \right)}^{-1} \left(\partial^i \partial^j V (\bar\p^i \s_a \p^j)+\frac 12 \partial^i V \partial^i V (\bar\chi^A \s_a \chi^A)\right) \mathcal{L}_a
\right.
\nonumber\\[2pt]
&&
\left.
\quad
+\left(\partial^i W^{jkl}-{ \left(\partial^p V \partial^p V \right)}^{-1} \left[\frac 12  \partial^i \partial^j V \partial^k \partial^l V +\partial^i \partial^l V \partial^k \partial^j V \right] \right) (\bar\p^i\p^j) (\bar\p^k \p^l)\right).
\eea
When shuffling between the Lagrangian and Hamiltonian formulations, it proves helpful to use the identity
\be
J_a={\left(\partial^l V \partial^l V \right)}^{-1} \left( \mathcal{L}_a- \left(\partial^i \partial^j V (\bar\p^i \s_a \p^j)+\frac 12 \partial^i V \partial^i V (\bar\chi^A \s_a \chi^A)\right)  \right),
\ee
which relates $J_a$ in (\ref{s3}) and $\mathcal{L}_a$ in (\ref{compL}). A possible alternative interpretation of the action (\ref{action1}) is that it describes a coupling of $M+1$ copies of $(4,4,0)$--supermultiplet, in which angular degrees of freedom are identified, to $N$ supermultiplets of the type $(0,4,4)$.

\vspace{0.5cm}

\noindent
{\bf 4. $D(2,1;\alpha)$ superparticles on near horizon black hole backgrounds}\\

By applying an appropriate canonical transformation, $(3,4,1)$--supermultiplet of the supergroup $SU(1,1|2)$ can be linked to the model of a massive relativistic superparticle propagating near the horizon of the extreme Reissner-Nordstr\"om black hole carrying the electric charge \cite{IKN,BGIK} or both the electric and magnetic charges \cite{G1}. Likewise, the near horizon geometry of the $d=5$, $N=2$ supergravity interacting with one vector multiplet turns out to be a proper background in the case of $(4,4,0)$--supermultiplet of $SU(1,1|2)$ \cite{G}. The two coordinate systems are referred to as the conformal and AdS bases \cite{IKN}.
In this section, we generalize the previous studies in \cite{G,BGIK,G1} to the case of the exceptional superconformal group $D(2,1;\alpha)$ and link the parameter $\alpha$ to the cosmological constant.

Consider the canonical transformation
\bea\label{CT}
&&
x'={ \left[\frac{2M^2}{x} \left(\sqrt{b^2+{(x p)}^2 +\alpha^2 J_a J_a}-b \right)\right]}^{\frac 12}\ ,
\nonumber\\[2pt]
&&
p'=-2 xp { \left[\frac{2M^2}{x} \left(\sqrt{b^2+{(x p)}^2 +\alpha^2 J_a J_a} -b \right)\right]}^{-\frac 12},
\nonumber\\[2pt]
&&
J'_a=J_a, \quad \psi'_\alpha=\psi_\alpha,
\eea
where $M$ and $b$ are real constants, $\alpha$ is the parameter entering $D(2,1;\alpha)$, and the prime designates coordinates in the conformal basis. Being rewritten in the AdS basis, the phase space functions (\ref{rep}) read
\bea
&&
H=\frac{x}{M^2}\left(\sqrt{b^2+{(x p)}^2 +\alpha^2 J_a J_a} +b
\right)
\nonumber\\[2pt]
&&
\qquad
+\frac{x}{M^2}
\left(\alpha (\bar\psi \sigma_a \psi) J_a-\frac 18 (1+2\alpha) \psi^2 \bar\psi^2\right) {\left(\sqrt{b^2+{(x p)}^2 +\alpha^2 J_a J_a}-b\right)}^{-1},
\nonumber\\[2pt]
&&
D=tH+x p, \qquad
K=t^2 H+2t x p+\frac{M^2}{x} \left(\sqrt{b^2+{(x p)}^2 +\alpha^2 J_a J_a}-b\right),
\nonumber
\eea
\bea\label{new}
&&
S_\a=\p_\a { \left(\frac{2M^2}{x} \left(\sqrt{b^2+{(x p)}^2 +\alpha^2 J_a J_a} -b \right)\right)}^{\frac 12}
-t Q_\a,
\nonumber\\[2pt]
&&
{\bar S}^\a={\bar\p}^\a { \left(\frac{2M^2}{x} \left(\sqrt{b^2+{(x p)}^2 +\alpha^2 J_a J_a} -b \right)\right)}^{\frac 12}
-t {\bar Q}^\a,
\nonumber\\[2pt]
&&
Q_\a=-\frac{2\left((x p) \p_\a+i \alpha{(\s_a \p)}_\a J_a +\frac i4 (1+2\alpha) \bar\p_\a \p^2 \right) }
{{ \left(\frac{2M^2}{x} \left(\sqrt{b^2+{(x p)}^2 +\alpha^2 J_a J_a} -b \right)\right)}^{\frac 12}}\ ,
\nonumber\\[2pt]
&&
{\bar Q}^\a=-\frac{2\left((x p) {\bar\p}^\a-i \alpha {(\bar\p \s_a)}^\a J_a +\frac i4 (1+2\alpha) \p^\a {\bar\p}^2 \right) }
{{ \left(\frac{2M^2}{x} \left(\sqrt{b^2+{(x p)}^2 +\alpha^2 J_a J_a} -b \right)\right)}^{\frac 12}}\ ,
\nonumber\\[2pt]
&&
\mathcal{J}_a=J_a+\frac 12 (\bar\p \s_a \p), \qquad I_{-}=\frac{i}{2} \psi^2, \qquad  I_{+}=-\frac{i}{2} {\bar\psi}^2, \qquad I_3=\frac 12 \bar\psi \psi.
\eea
Because the transformation (\ref{CT}) is canonical, they do obey the structure relations of the Lie superalgebra corresponding to $D(2,1;\alpha)$.

The important point regarding the realization (\ref{new}) is that
omitting the fermions one obtains a bosonic system whose structure is typical for a massive relativistic particle propagating in a curved spacetime. Let us identify backgrounds corresponding to
$D(2,1;\alpha)$ superconformal mechanics based on three realizations of $su(2)$ in Eqs. (\ref{S2}), (\ref{s2}), (\ref{s3}) above.

As the first step, consider the metric and the gauge field one--form
\bea\label{RN}
&&
ds^2={\left(\frac{r}{M} \right)}^2 dt^2-  {\left(\frac{M}{r} \right)}^2 dr^2-{\left(\frac{M}{\a} \right)}^2 (d \theta^2+\sin^2{\theta} d \phi^2),
\nonumber\\[2pt]
&&
A=\frac{Q}{M^2} r dt+P \cos{\theta} d \phi,
\eea
where $M$, $Q$, $P$, and $\alpha$ are constants. One can readily verify that these fields do obey the Einstein--Maxwell equations with the cosmological term
\be
R_{nm}-\frac 12 g_{nm} (R+2 \Lambda)=-2(F_{ns} {F_m}^s-\frac 14 g_{nm} F^2), \qquad  \partial_n (\sqrt{-g} F^{nm})=0,
\ee
provided the conditions
\be\label{COND}
M = \sqrt{\frac{2 (Q^2+\a^4 P^2)}{1 + \a^2}},
\qquad \Lambda=\frac{\a^2-1}{2 M^2 }
\ee
hold. Eq. (\ref{RN}) describes the near horizon geometry of the extreme Reissner--Nordstr\"om-AdS-dS black hole,
$M$, $Q$, and $P$ being the mass, the electric and magnetic charges, respectively. Remarkably enough, the parameter $\alpha$ is linked to the cosmological constant. In particular, $\alpha^2=1$ yields $\Lambda=0$, while the domains $\alpha^2<1$ and $\alpha^2>1$ correspond to the negative and positive cosmological constants. The value $\alpha=0$ is excluded from the consideration as the metric becomes singular.

Then let us demonstrate that $D(2,1;\alpha)$ superconformal mechanics based on the realizations of $su(2)$ in (\ref{S2}), (\ref{s2}) can be linked to the superparticle propagating near the horizon of the extreme Reissner--Nordstr\"om-AdS-dS black hole in four dimensions. Consider the static gauge action functional of a massive particle coupled to the background fields (\ref{RN})
\bea\label{start}
&&
S=-\int \left(m ds+e A \right)=-\int d t \left(m\sqrt{ {(r/M)}^2-{(M/r)}^2 {\dot r}^2 -{(M/\a)}^2 ({\dot\t}^2
+\sin^2 \t {\dot\phi}^2)   }
\right.
\nonumber\\[2pt]
&&
\left.
\qquad \qquad \qquad \qquad  \qquad
+eQr/M^2 +e P \cos \t \dot\phi~\right),
\eea
where $m$ and $e$ designate its mass and electric charge. Introducing momenta $(p_r,p_\t,p_\phi)$ canonically
conjugate to the configuration space variables $(r,\t,\phi)$, one can readily construct the Hamiltonian
\be\label{h}
H=\frac{r}{M^2} \left(\sqrt{ {(m M)}^2+{(r p_r)}^2 +\alpha^{2} \left(p_\t^2+\sin^{-2}\t {(p_\phi+e P \cos\t)}^2 \right)} +e Q \right).
\ee
Taking into account the last line in Eq. (\ref{s2}) and the first line in Eq. (\ref{new}), one concludes that the model (\ref{h}) is amenable to $D(2,1;\alpha)$ superconformal extension provided the BPS--like condition is imposed on the particle parameters
\be
{(e Q)}^2={(m M)}^2-{(\alpha e P)}^2.
\ee
For $\alpha^2=1$ the latter correctly reproduces the analysis in \cite{BGIK,G1}.
The two (on--shell) versions of $(3,4,1)$--supermultiplet associated with the realizations of $su(2)$  in (\ref{S2}) and (\ref{s2}) can thus be linked to a $D(2,1;\alpha)$ superparticle propagating near the horizon of the extreme Reissner--Nordstr\"om-AdS-dS black hole which carries either electric or both the electric and magnetic charges.

Finally, let us identify background geometry which can be connected to $(4,4,0)$--supermul\-tiplet of $D(2,1;\alpha)$ based on Eq. (\ref{s3}).
Consider the equations of motion which describe the bosonic limit of the $d=5$, $N=2$ supergravity interacting with one vector multiplet\footnote{Our notations are similar to those in \cite{cfgk}.
We use the mostly minus signature convention for the metric $g_{mn}$ and set $g=\det{g_{mn}}$.}
in spacetime with cosmological constant
\bea\label{EqM}
&&
R_{mn}-\frac 12 g_{mn} (R+2 \Lambda)+e^{\frac 23 \varphi }  (F_{mk} {F_n}^k-\frac 14 g_{mn} F^2)+e^{-\frac 43 \varphi }  (G_{mk} {G_n}^k-\frac 14 g_{mn} G^2)
\nonumber\\[2pt]
&&
-\frac 13 (\partial_m \varphi \partial_n \varphi-\frac 12 g_{mn} \partial_k \varphi \partial^k \varphi)=0, \quad
\nabla_m \left(e^{\frac 23 \varphi } F^{mn}-\frac{1}{\sqrt{2g}} \epsilon^{mnpqr} F_{pq} B_r \right)=0,
\nonumber\\[2pt]
&&
\nabla_m \left(e^{-\frac 43 \varphi } G^{mr} \right)+\frac{1}{4 \sqrt{2 g}} \epsilon^{mnpqr} F_{mn} F_{pq} =0, \qquad \nabla^2 \varphi+\frac 12 e^{\frac 23 \varphi } F^2-e^{-\frac 43 \varphi } G^2=0,
\eea
where $\varphi$ is a scalar field, and $F_{nm}=\partial_n A_m-\partial_m A_n$, $G_{nm}=\partial_n B_m-\partial_m B_n$, $F^2=F_{nm} F^{nm}$, $G^2=G_{nm} G^{nm}$.
It is straightforward to verify that the set of fields
\bea\label{backfields}
&&
ds^2
={\left(\frac{r}{M}\right)}^2 dt^2-{\left(\frac{M}{r}\right)}^2 dr^2-{\left( \frac{M}{\alpha} \right)}^2 \left(d \theta^2+\sin^2{\theta} d \phi^2+{(d \psi+\cos{\theta} d \phi)}^2\right),
\nonumber\\[2pt]
&&
A=\frac{Q r}{M} dt, \qquad B=\frac{Q r}{\sqrt{2} M} dt, \qquad \varphi=0,
\eea
where $M$, $Q$, and $\alpha$ are constants, does solve (\ref{EqM}) provided the constraints
\be
\Lambda=\frac{\alpha^2-1}{2M^2}, \qquad Q=\pm \sqrt{\frac{2+\alpha^2}{3}}
\ee
hold.

Because a charged massive particle couples only to the electromagnetic field and gravity, the static gauge action functional reads
\bea
&&
S=
-\int d t \left(m\sqrt{ {(r/M)}^2-{(M/r)}^2 {\dot r}^2 -{(M/\a)}^2 \left({\dot\t}^2
+\sin^2 \t {\dot\phi}^2+{(\dot\psi+\cos{\theta} \dot\phi)}^2\right)   }
\right.
\nonumber\\[2pt]
&&
\qquad
\left.
+eQr/M \right),
\eea
where $m$ and $e$ are the mass and electric charge of a particle probe. The corresponding canonical Hamiltonian takes the form
\be\label{HAM}
H=\frac{r}{M^2} \left(\sqrt{ {(m M)}^2+{(r p_r)}^2 +\alpha^{2} \left(p_\t^2+\sin^{-2}\t {(p_\phi-p_\psi \cos\t)}^2 +p_\psi^2\right)} +e Q M \right),
\ee
where $(p_\theta,p_\phi,p_\psi)$ denote momenta canonically conjugate to $(\theta,\phi,\psi)$. As follows from the last line in Eq. (\ref{s2}) and the first line in Eq. (\ref{new}),
the model (\ref{HAM}) is amenable to $D(2,1;\alpha)$ superconformal extension provided the BPS--like condition on the particle parameters
\be
m^2=\frac{(2+\alpha^2) e^2}{3}
\ee
holds.

We thus conclude that $(4,4,0)$--supermultiplet of $D(2,1;\alpha)$ based on the realization of $su(2)$ in (\ref{s3}) can be linked to a near horizon BPS--superparticle minimally coupled to fields of the $d=5$, $N=2$ supergravity interacting with one vector multiplet in spacetime with cosmological constant. As in the preceding case, the parameter $\alpha$ turns out to be related to the cosmological constant.

\vspace{0.5cm}

\noindent
{\bf 5. Conclusion}

\vspace{0.5cm}

To summarize, in this work we generalized the analysis in \cite{G} to the case of the most general $N=4$ superconformal group in one dimension $D(2,1;\alpha)$. It was shown that
any realization of the $R$--symmetry subalgebra $su(2)$ in terms of phase space functions can be extended to a representation of the Lie superalgebra corresponding to $D(2,1;\alpha)$.
Novel coupling of arbitrary number of supermultiplets of the type $(1,4,3)$ and $(0,4,4)$ to a single supermultiplet of either the type $(3,4,1)$, or $(4,4,0)$ has been constructed
by arranging the $su(2)$--generators so as to include both bosons and fermions. Alternatively, this system can be viewed as describing an interaction of $M+1$ copies of either $(3,4,1)$--, or $(4,4,0)$--supermultiplet, in which angular degrees of freedom are identified, with $N$ supermultiplets of the type $(0,4,4)$. A canonical transformation which relates $D(2,1;\alpha)$ superconformal mechanics based on supermultiplets of the type
$(3,4,1)$ and $(4,4,0)$ to BPS--superparticles propagating near the horizon of the extreme Reissner--Nordstr\"om--AdS--dS black hole in four and five dimensions was found.
The parameter $\alpha$ was linked to the cosmological constant.

There are several directions in which the present work can be continued. First of all, it would be interesting to construct an off--shell superfield Lagrangian formulation for the component Hamiltonian framework presented in Sect. 3.
Interacting systems in Sect. 3 were interpreted as describing a coupling of arbitrary number of supermultiplets of the type $(1,4,3)$ and $(0,4,4)$ to a single supermultiplet of either the type $(3,4,1)$, or $(4,4,0)$. As was mentioned above, in principle, an alternative interpretation is possible in which several copies of $(3,4,1)$--, or $(4,4,0)$--supermultiplets are first identified along angular degrees of freedom and then they are coupled to $(0,4,4)$--supermultiplets. It is interesting to understand whether superfield constrains leading to such an identification along the angular degrees of freedom can be formulated in superspace.
A $\kappa$--symmetric Lagrangian formulation for the BPS--superparticles in Sect. 4 and a possible connection between the supersymmetry charges and the Killing spinors characterizing the background geometry are worth studying as well. Finally, it is of interest to study the models in this work from the perspective of the Kirillov--Kostant--Souriau method (see a recent work \cite{HJM} and references therein).

\vspace{0.5cm}

\noindent{\bf Acknowledgements}\\

\noindent
We thank E. Ivanov for a useful correspondence.

\noindent

\vspace{0.5cm}

\noindent
{\bf Appendix}

\vspace{0.5cm}
\noindent
Throughout the text $SU(2)$--spinor
indices are raised and lowered with the use of the invariant
antisymmetric matrices
\be
\p^\a=\e^{\a\b}\p_\b, \quad {\bar\p}_\a=\e_{\a\b} {\bar\p}^\b,
\nonumber
\ee
where $\e_{12}=1$, $\e^{12}=-1$. Introducing the notation for the spinor bilinears
\be
\p^2=(\p^\a \p_\a\ ), \qquad
\bar\p^2=(\bar\p_\a \bar\p^\a ), \qquad \bar\p \p=(\bar\p^\a \p_\a ),
\nonumber
\ee
one gets
\be
\p_\a \p_\b=\frac 12 \e_{\a\b} \p^2, \qquad \p_\a \bar\chi_\b-\p_\b \bar\chi_\a=\e_{\a\b} (\bar\chi \p),
\nonumber
\ee
\be
\bar\p^\a \bar\p^\b=\frac 12 \e^{\a\b} \bar\p^2,  \qquad \psi^\alpha {\bar\chi}^\beta-\psi^\beta {\bar\chi}^\alpha=-\epsilon^{\alpha \beta} (\bar\chi\psi).
\nonumber
\ee
The Pauli matrices ${{(\s_a)}_\a}^\b$
are chosen in the standard form
\be
\s_1=\begin{pmatrix}0 & 1\\
1 & 0
\end{pmatrix}\ , \qquad \s_2=\begin{pmatrix}0 & -i\\
i & 0
\end{pmatrix}\ ,\qquad
\s_3=\begin{pmatrix}1 & 0\\
0 & -1
\end{pmatrix}\ ,
\nonumber
\ee
which obey
\bea
&&
{{(\s_a \s_b)}_\a}^\b +{{(\s_b \s_a)}_\a}^\b=2 \d_{ab} {\d_\a}^\b \ , \quad
{{(\s_a \s_b)}_\a}^\b -{{(\s_b \s_a)}_\a}^\b=2i \e_{abc} {{(\s_c)}_\a}^\b \ ,
\nonumber\\[2pt]
&&
{{(\s_a \s_b)}_\a}^\b=\d_{ab} {\d_\a}^\b +i \e_{abc} {{(\s_c)}_\a}^\b \ , \quad
{{(\s_a)}_\a}^\b {{(\s_a)}_\g}^\r=2 {\d_\a}^\r {\d_\g}^\b-{\d_\a}^\b {\d_\g}^\r\ ,
\nonumber\\[2pt]
&&
{{(\s_a)}_\a}^\b \e_{\b\g} ={{(\s_a)}_\g}^\b \e_{\b\a}\ , \quad \e^{\a\b} {{(\s_a)}_\b}^\g=\e^{\g\b} {{(\s_a)}_\b}^\a \ ,
\nonumber
\eea
where $\e_{abc}$ is the totally antisymmetric tensor, $\e_{123}=1$. Throughout the text we denote
$\bar\p \s_a \p=\bar\p^\a {{(\s_a)}_\a}^\b \p_\b$. Our conventions for complex conjugation read
\bea
&&
{(\psi_\alpha)}^{*}=\bar\psi^\alpha, \qquad {(\bar\psi_\alpha)}^{*}=-\psi^\alpha, \qquad
{(\psi^2)}^{*}=\bar\psi^2, \qquad {(\bar\psi \sigma_a \chi)}^{*}=\bar\chi \sigma_a \psi.
\nonumber
\eea

\end{document}